\input harvmac.tex
\hfuzz 15pt
\input amssym.def
\input amssym.tex
\input epsf\def\tfig#1{{
\xdef#1{Fig.\thinspace\the\figno}}Fig.\thinspace\the\figno
\global\advance\figno by1}


\input epsf

%



\def\p{\partial}

\def\a{\alpha}
\def\b{\beta}
\def\g{\gamma}

\def\e{\epsilon}

\def\k{\kappa}
\def\l{\lambda}

\def\G{\Gamma}
\def\D{\Delta}

\def\ph{\phi}
\def\ov{\over}

\def\no{\noindent}


 %

\def\[{\left[}
\def\]{\right]}
\def\({\left(}
\def\){\right)}
\def\<{\left\langle\,}
\def\>{\,\right\rangle}


\def\inv{^{-1}}

\def\hf{{\textstyle{1\over 2}}}

 \def\frac#1#2{ {{\textstyle{#1\over#2}}}}
\def\inv{^{\raise.15ex\hbox{${\scriptscriptstyle -}$}\kern-.05em 1}}

 \def\IP{\relax{\rm I\kern-.18em P}}


%




\def\dC{C\kern-6.5pt I}

       \def\CB{{\cal B}}       \def\CC{{\cal C}}
              
\def\CG{{\cal G}}              
              \def\CL{{\cal L}}

       \def\CT{{\cal T}}

%



\chardef\tempcat=\the\catcode`\@ \catcode`\@=11
\def\cyracc{\def\u##1{\if \i##1\accent"24 i%
    \else \accent"24 ##1\fi }}
\newfam\cyrfam



\def\np#1#2#3{{Nucl. Phys.} {\bf B#1} (#2) #3}
\def\pl#1#2#3{{Phys. Lett. }{\bf B#1} (#2) #3}

\def\physrev#1#2#3{{Phys. Rev.} {\bf D#1} (#2) #3}

\def\cmp#1#2#3{{Comm. Math. Phys.} {\bf #1} (#2) #3}
\def\mpl#1#2#3{{Mod. Phys. Lett. }{\bf #1} (#2) #3}
\def\ijmp#1#2#3{{Int. J. Mod. Phys.} {\bf #1} (#2) #3}
\def\lmp#1#2#3{{Lett. Math. Phys.} {\bf #1} (#2) #3}
\def\tmatp#1#2#3{{Theor. Math. Phys.} {\bf #1} (#2) #3}
\def\hep#1{{hep-th/}#1}

\def\encadremath#1{\vbox{\hrule\hbox{\vrule\kern8pt\vbox{\kern8pt
 \hbox{$\displaystyle #1$}\kern8pt}
 \kern8pt\vrule}\hrule}}

\def\tphi{\tilde\phi}


\def\hepth#1{{arXiv:hep-th/}#1}

\def\np#1#2#3{{Nucl. Phys.} {\bf B#1} (#2) #3}
\def\pl#1#2#3{{Phys. Lett. }{\bf B#1} (#2) #3}

\def\physrev#1#2#3{{Phys. Rev.} {\bf D#1} (#2) #3}

\def\cmp#1#2#3{{Comm. Math. Phys.} {\bf #1} (#2) #3}
\def\mpl#1#2#3{{Mod. Phys. Lett. }{\bf #1} (#2) #3}
\def\ijmp#1#2#3{{Int. J. Mod. Phys.} {\bf #1} (#2) #3}
\def\lmp#1#2#3{{Lett. Math. Phys.} {\bf #1} (#2) #3}
\def\tmatp#1#2#3{{Theor. Math. Phys.} {\bf #1} (#2) #3}
\def\jhep#1#2#3{{JHEP} {\bf #1} (#2) #3}




\lref\gko{P. Goddard, A. Kent, D. Olive, Virasoro algebras and coset space models, \pl{152}{1985}88.}
\lref\myo{C. Crnkovic, G. Sotkov, M. Stanishkov, Renormalization group flow for general $SU(2)$ coset models, \pl{226}{1989}297.}
\lref\rav{F. Ravanini, Thermodynamic Bethe ansatz for G(k) x G(l) / G(k+l) coset models perturbed by their phi(1,1,Adj) operator, \pl{282}{1992}73.}
\lref\zam{A. Zamolodchikov, Renormalization Group and Perturbation Theory Near Fixed Points in Two Dimensional Field Theory, Sov. J. Nucl. Phys. {\bf 46} (1987) 1090.}
\lref\pogo{R. Poghossian, Study of the Vicinities of Superconformal Fixed Points in Two-dimensional Field Theory, Sov. J. Nucl. Phys. {\bf 48} (1988) 763.}
\lref\pogt{R. Poghossian, Two Dimensional Renormalization Group Flows in Next to Leading Order, \jhep{1401}{2014}167; \hepth{1303.3015}.}
\lref\mymy{C. Ahn, M. Stanishkov, On the Renormalization Group Flow in Two Dimensional Superconformal Models, \np{885}{2014}713; \hepth{1404.7628}.}
\lref\myt{C. Crnkovic, R. Paunov, G. Sotkov, M. Stanishkov, Fusions of Conformal Models, \np{336}{1990}637.}
\lref\gai{D. Gaiotto, Domain Walls for Two-Dimensional Renormalization Group Flows, \jhep{1212}{2012}103; \hepth{1201.0767}.}
\lref\ibr{I. Brunner, D. Roggenkamp, Defects and bulk perturbations of boundary Landau-Ginzburg orbifolds, \jhep{0804}{2008}001; \hepth{0712.0188}.}
\lref\marco{F. Gliozzi, P. Liendo, M. Meineri, A. Rago, Boundary and Interface CFTs from the Conformal Bootstrap, \jhep{1505}{2015}036; \hepth{1502.07217}.}
\lref\ppo{A. Poghosyan, H. Poghosyan, Mixing with descendant fields in perturbed minimal CFT models, \jhep{1310}{2013}131; \hepth{1305.6066}.}
\lref\ppt{A. Poghosyan, H. Poghosyan, RG domain wall for the $N=1$ minimal superconformal models, \jhep{1505}{2015}043; \hepth{1412.6710}.}
\lref\ibc{I.Brunner, C. Schmidt-Colinet, Reflection and transmission of conformal perturbation defects, J.Phys. {\bf A49}(2016)195401; \hepth{1508.04350}.}
\lref\kmq{D. Kastor, E. Martinec, Z. Qiu, Current Algebra and Conformal Discrete Series, \pl{200}{1988}434.}
\lref\fr{F. Ravanini, An Infinite Class of New Conformal Field Theories With Extended Algebras, \mpl{3A}{1988}397.}
\lref\agt{P. Argyres, J. Grochocinski, S. Tye, Structure Constants of the Fractional Supersymmetry Chiral Algebras, \np{367}{1991}217; \hepth{9110052}.}
\lref\kt{Z. Kakuushadze, S.Tye, Kac and new determinants for fractional superconformal algebras, \physrev{49}{1994}4122; \hepth{9310160}.}
\lref\willa{N. Wyllard, Coset conformal blocks and N=2 gauge theories, \hepth{1109.4264}.}
\lref\df{V. Dotsenko, V. Fateev, Operator Algebra of Two Dimensional Conformal Theories with Central  Charge $c<1$, \pl{154}{1985}291}
\lref\zp{A. Zamolodchikov, R. Poghossian,  Operator algebra in two dimensional superconformal field theory, Sov. J. Nucl. Phys. {\bf 47}(1988)929.}
\lref\pogtri{R. Poghossian, Operator Algebra in Two Dimensional Conformal Quantum Field Theory Containing Spin $4/3$ Parafermionic Coserved Currents, \ijmp{A6}{1991}2005.}

\overfullrule=0pt
\Title{\vbox{\baselineskip12pt\hbox {}\hbox{}}} {\vbox{\centerline
 {RG Domain Wall for the}
  \vskip10pt
\centerline{General $\hat{su}(2)$ Coset Models }
  \vskip2pt
}} \centerline{ Marian Stanishkov\foot{marian@inrne.bas.bg } }

 \vskip 1cm

 \centerline{ \vbox{\baselineskip12pt\hbox {\it Institute for
Nuclear Research and Nuclear Energy,}
 }}
\centerline{ \vbox{\baselineskip12pt\hbox {\it Bulgarian Academy of Sciences, 1784 Sofia, Bulgaria}
 }}


\vskip 1.5cm

\centerline{ Abstract} \vskip.5cm \noindent \vbox{\baselineskip=11pt
We consider a RG flow in a general $\hat{su}(2)$ coset model induced by the least relevant field. This is done using two different approaches. We first compute the mixing coefficients of certain fields in the UV and IR theories using a conformal perturbation theory. The necessary structure constants are computed. The same coefficients can be calculated using the RG domain wall construction of Gaiotto. We compute the corresponding one-point functions and show that the two approaches give the same result in the leading order. }

\Date{}
\vfill \eject

%

\newsec {Introduction}

In this paper we consider the general $\hat{su}(2)$ coset model $M(k,l)$ \gko\ perturbed by the least relevant operator. This theory was already analyzed in \myo\ (see also \rav\ ). It was shown there that in the leading order in $1/k$, $k\rightarrow\infty$, there exists a nontrivial fixed point, a zero of the $\b$-function. The IR theory was identified with $M(k-l,l)$.  Here we are interested in the mixing of certain fields under the corresponding RG flow. This is interesting problem because, as it can be seen from \zam\ and \pogo\ in the first order, and in \pogt\ and \mymy\ in the second, the mixing coefficients are the same for $l=1$ (Virasoro) and $l=2$ (superconformal) theories. We will show that this is the case in the general theory, i.e.  they do not depend on $l$ and are finite in the leading order. We find it convenient, following \myo , to use the construction presented in \myt . Namely, we define the perturbing field and the other fields in consideration recursively as a product of lower level fields. Then the corresponding structure constants, governing the perturbation expansion, are obtained by demanding the closing of the OPE's with the perturbing field.

In the second part of the paper we present an alternative, non-perturbative description of the RG flow. It was proposed time ago by Gaiotto in \gai\ where he specified the construction of \ibr\ for the case of perturbed minimal models. In this approach the UV-IR map resulting through the RG flow is encoded in a specific conformal interface, the RG domain wall. In this construction the mixing coefficients are expressed in terms of the one point functions of the product of UV and IR theories in the presence of a specific RG boundary.\foot{In \marco\ it is shown that this is the case also for other models and for any dimension.}
The corresponding boundary state was constructed in \gai . It was shown there that for $l=1$ the coefficients coincide with those found in \zam\ in the leading order. The same is true also in the second order \pogt\foot
 {As explained in \pogt\ this is valid up to eventual reformulation of the renormalization scheme.}.The construction was further generalized for descendent fields \ppo\ in $l=1$ and for some fields in $l=2$ \ppt\ where also the coefficients coincide. The case of general level $l$ theories was addressed in \ibc\ where reflection and transmission coefficients were found. In this paper we show that the mixing coefficients of certain fields are the same for any $l$ in the leading order.

This paper is organized as follows. In the next section we describe the perturbation of the general $\hat{su}(2)$ coset theory. We construct the perturbing field as well as special fields that mix under the corresponding RG flow. The structure constants of the latter are computed following an approach analogous to \myo . In the third section we use the structure constants to compute the mixing coefficients of these fields in the leading order. Section 4 is devoted to the construction of the domain wall for the perturbed coset model. We compute the one-point functions corresponding to the mixing coefficients. They turn out to coincide with those of section 3 in the leading order. Finally, we present our conclusions.

\newsec{The theory}

Consider a two-dimensional CFT $M(k,l)$ based on the coset
\eqn\coset{
{\hat{su}(2)_k\times \hat{su}(2)_l\over \hat{su}(2)_{k+l}}}
where $k$ and $l$ are integers and we assume here that $k>l$.
It is written in terms of $\hat{su}(2)_k$ WZNW models with current $J^a$, $k$ is the level. The latter are CFT's with a stress tensor expressed through the currents by the Sugawara construction:
\eqn\suga{
T_k(z)={1\ov k+2}\((J^0)^2+\hf J^+J^-+\hf J^-J^+\).}
The central charge of the corresponding Virasoro algebra is
$c_k={3k\ov k+2}$.
The energy momentum tensor of the coset \coset\ is then given by: $T=T_k+T_l-T_{k+l}$ in obvious notations.
The resulting central charge for the coset CFT can be read from this construction:
$$
c={3kl(k+l+4)\over (k+2)(l+2)(k+l+2)}={3l\over l+2}\(1-{2(l+2)\over (k+2)(k+l+2)}\).
$$
The dimensions of the primary fields $\phi_{m,n}(l,p)$ of the "minimal models" (rational CFT) are given by \kmq\ ($m,n$ are integers) :
\eqn\dim{
\eqalign{
\D_{m,n}(l,p) &={((p+l)m-p n)^2-l^2\over 4lp(p+l)}+{s(l-s)\over 2l(l+2)},\cr
s &=|m-n|( mod (l)),\hskip1cm 0\le s\le l,\cr
&1\le m\le p-1, \hskip1cm 1\le n\le p+l-1}}
where we introduced ${\bf p=k+2}$ (note that we inverted $k$ and $l$  in the definition of the fields since we want to follow the notations of \myo\ ).
It is known \refs{\kmq ,\fr ,\agt} that the theory $M(k,l)$ possesses a symmetry generated by a "parafermionic current" $A(z)$ of dimension $\D_A={l+4\over l+2}$.
We shall present an explicit construction of this current below. Here we just mention that under this symmetry
the primary fields \dim\ are divided in sectors labeled by the integer $s$.
The branching of the current $A(z)$ on the field (or state) of sector $s$ can be written symbolically as \kt\ (see also \willa\ ):
\eqn\branch{
A_{-m-{(s+2)\over (l+2)}}|s>=|s+2>,\hskip.5cm A_{-m}|s>=|s>,\hskip.5cm A_{-m-{(l+2-s)\over (l+2)}}|s>=|s-2>.}

In this paper we prefer to use another description of the theory
$M(k,l)$ presented in \myt . It was shown there that this theory is not
independent but can be built out of a product of theories of lower
level. Schematically this can be written as a recursion:
$$
M(1,l-1)\times M(k,l)={\bf P}(M(k,1)\times M(k+1,l-1))
$$
where ${\bf P}$ in the RHS is a specific projection. It allows the
multiplication of fields of the same internal indices and describes
primary and descendent fields (see \myt\ for more details).

In this paper we consider the CFT $M(k,l)$ perturbed by the least
relevant field. Such theory was described in \myo\ where the
$\b$-function and the fixed point were found. Our goal here
is to describe also the mixing of certain fields under the RG flow.

Let us briefly sketch the constructions. The perturbed theory is
described by the Lagrangian:
$$
\CL(x)=\CL
_0(x)+\l \tilde\phi_{1,3}(x)
$$
where $\CL_0(x)$ describes the theory $M(k,l)$ itself. We
identify the field $\tilde\phi_{1,3}$ with the first descendent of
the corresponding primary field \dim\ with respect to the current
$A(z)$. In fact, in view of \dim\ $\phi_{1,3}$ belongs to the sector
$|2>$ and has a descendent belonging to sector $|0>$ due to the last
of \branch . The dimension of this first descendent is therefore
(for $s=2$):
\eqn\delt{
\D=\D_{1,3}+{l\over l+2}=1-{2\over p+l}=1-\e.}
In this and in the next sections we consider the case $p\rightarrow\infty$ and
assume that $\e={2\over p+l}\ll 1$ is a small parameter.

Following \myo\ we find it more convenient here to define the field $\tilde\phi_{1,3}$
alternatively in terms of lower level fields:
\eqn\field{
\tilde\phi_{1,3}(l,p)=a(l,p)\phi_{1,1}(1,p)\tilde\phi_{1,3}(l-1,p+1)+b(l,p)\phi_{1,3}(1,p)\phi_{3,3}(l-1,p+1)}
where
$a=\sqrt{{(l-1)(p-2)\ov l(p-1)}},\quad b=\sqrt{{p-l-2\ov l(p-1)}}.$

The field $\phi_{3,3}(l,p)$ is just a primary field constructed as \myo\ :
\eqn\fitri{
\phi_{3,3}(l,p)=\phi_{3,3}(1,p)\phi_{3,3}(l-1,p+1)}
with dimension from \dim. It is straightforward to check that the field \field has a correct dimension \delt.

The mixing of the fields along the RG flow is connected to the two-point function. In the first order of the perturbation theory it is given by:
\eqn\pert{
<\ph_\a(x_1)\ph_\b(x_2)>=<\ph_\a(x_1)\ph_\b(x_2)>_0-\l \int <\ph_\a(x_1)\ph_\b(x_2)\tilde\phi_{1,3}(x)>_0 d^2x+O(\l^2)}
The integral in \pert\ is governed by the corresponding structure constants:
\eqn\firstint{
\eqalign{C_{\a,\b}^{(1)}&=\int <\phi_\a(1)\phi_\b(0)\tilde\phi_{1,3}(x)>d^2x=\cr
&=C_{(1,3)(\a)(\b)}{\pi\g(\e+\D_\a-\D_\b)\g(\e+\D_\b-\D_\a)\ov \g(2\e)}}}
Therefore the computation of the structure constants is of primary importance.

The coefficients $a(l,p)$ and $b(l,p)$, as well as the necessary structure constants, were obtained in \myo\ .

For example the structure constant we need for the computation of the $\b$-function (up to the desired order) is:
$$
C_{(13)(13)}^{(13)}(l,p))={4\ov l\sqrt{3}}+O(\e).
$$
The computation of the $\b$-function gives:
$$
\b=\e g-{1\ov 2}\pi C_{(13)(13)}^{(13)}(l,p)g^2+...=\e g-{2\pi\ov l\sqrt{3}}g^2+...
$$
There is an obvious non trivial fixed point:
\eqn\fx{
g^*={\sqrt{3} l\ov 2\pi}\e+...}
It was shown in \myo\ that this fixed point corresponds to the theory $M(k-l,l)$. Indeed:
$$
c^*-c_p=-{(l+2)\ov 3l}12\pi^2\int_0^{g^*}
 \b(g)dg=-{l(l+2)\ov 2}\e^3+...
$$
which perfectly coincides with $c_{p-l}-c_p$ up to this order in $\e$.
The anomalous dimension of the field $\tilde\phi_{1,3}(l,p)$ at the fixed point is:
$$
\D^*=1-\p_g\b(g)|_{g=g^*}=1+\e+O(\e^2).
$$
This dimension matches up to this order the dimension of the field $\tilde\phi_{3,1}(l,p-l)$ which is defined analogously to $\tilde\phi_{1,3}(l,p)$. We shall present its exact construction below.

Let us define recursively, in analogy with the fields $\tilde\phi_{1,3}(l,p)$ and $\phi_{3,3}(l,p)$, the following descendent fields:
\eqn\defn{
\eqalign{
\tilde\phi_{n,n+2}(l,p)&=x(l,p)\phi_{n,n}(1,p)\tilde\phi_{n,n+2}(l-1,p+1)+\cr
&+y(l,p)\phi_{n,n+2}(1,p)\phi_{n+2,n+2}(l-1,p+1),\cr
\tilde\phi_{n,n-2}(l,p)&=\tilde x(l,p)\phi_{n,n}(1,p)\tilde\phi_{n,n-2}(l-1,p+1)+\cr
&+\tilde y(l,p)\phi_{n,n-2}(1,p)\phi_{n-2,n-2}(l-1,p+1)}}
and the primary field
\eqn\defnn{
\phi_{n,n}(l,p)=\phi_{n,n}(1,p)\phi_{n,n}(l-1,p+1).}
The dimensions of these fields are:
\eqn\dimn{
\eqalign{
\tilde\D_{n,n\pm 2} &=1+{n^2-1\ov 4p}-{(2\pm n)^2-1\ov 4(p+l)}=1-{1\pm n\ov 2}\e+O(\e^2),\cr
\D_{n,n} &={n^2-1\ov 4p}-{n^2-1\ov 4(p+l)}={(n^2-1)l\ov 16}\e^2+O(\e^3).}}
They are analogs of the (descendants of the) NS fields of the $N=1$ super conformal theory($l=2$) and the fields from $S$ or $D$-sectors of $4/3$-parafermionic theory ($l=4$).

Two remarks are in order. First, similarly to $\tilde\phi_{1,3}(l,p)$ and $\phi_{3,3}(l,p)$ the fields defined above belong to the zero charge, or "vacuum sector". The arguments for that go along the same lines. Second, the fields \defn\ and the derivative of \defnn\ have dimensions close to one and therefore can mix. To ensure this we ask that their fusion rules with the perturbing field are closed. This requirement defines the coefficients in \defn\ and the corresponding structure constants.
So we impose the conditions:
\eqn\frn{
\eqalign{
\tilde\phi_{1,3}(l,p)\tilde\phi_{n,n+2}(l,p) &=\CC_{(13)(nn+2)}^{(nn)}(l,p)\phi_{n,n}(l,p)+\CC_{(13)(nn+2)}^{(nn+2)}(l,p)\tilde\phi_{n,n+2}(l,p),\cr
\phi_{3,3}(l,p)\phi_{n,n}(l,p) &=\CC_{(33)(nn)}^{(nn+2)}(l,p)\tilde\phi_{n,n+2}(l,p)+\CC_{(33)(nn)}^{(nn)}(l,p)\phi_{n,n}(l,p).}}
As in \myo , using the constructions \field, \fitri\ and \defn, \defnn, we obtain functional equations for the coefficients and the structure constants:
\eqn\eone{
\eqalign{
&a x \CC_{(13)(nn+2)}^{(nn+2)}(l-1,p+1)+b x  \CC_{(13)(nn)}^{(nn)}(1,p)\CC_{(33)(nn+2)}^{(nn+2)}(l-1,p+1)+\cr
&+b y  \CC_{(13)(nn+2)}^{(nn)}(1,p)\CC_{(33)(n+2n+2)}^{(nn+2)}(l-1,p+1)=x \CC_{(13)(nn+2)}^{(nn+2)}(l,p),}}
\eqn\etwo{
\eqalign{
&a y \CC_{(13)(n+2n+2)}^{(n+2n+2)}(l-1,p+1)+b x  \CC_{(13)(nn)}^{(nn+2)}(1,p)\CC_{(33)(nn+2)}^{(n+2n+2)}(l-1,p+1)+\cr
&+b y  \CC_{(13)(nn+2)}^{(nn+2)}(1,p)\CC_{(33)(n+2n+2)}^{(n+2n+2)}(l-1,p+1)=y \CC_{(13)(nn+2)}^{(nn+2)}(l,p).}}
\eqn\etri{
\eqalign{
&a x \CC_{(13)(nn+2)}^{(nn)}(l-1,p+1)+b x  \CC_{(13)(nn)}^{(nn)}(1,p)\CC_{(33)(nn+2)}^{(nn)}(l-1,p+1)+\cr
&+b y  \CC_{(13)(nn+2)}^{(nn)}(1,p)\CC_{(33)(n+2n+2)}^{(nn)}(l-1,p+1)= \CC_{(13)(nn+2)}^{(nn)}(l,p)}}
from the first of \frn\ and
\eqn\efor{
\eqalign{
& \CC_{(33)(nn)}^{(nn)}(1,p)\CC_{(33)(nn)}^{(nn+2)}(l-1,p+1)=x \CC_{(33)(nn)}^{(nn+2)}(l,p),\cr
& \CC_{(33)(nn)}^{(nn+2)}(1,p)\CC_{(33)(nn)}^{(n+2n+2)}(l-1,p+1)=y \CC_{(33)(nn)}^{(nn+2)}(l,p),\cr
& \CC_{(33)(nn)}^{(nn)}(1,p)\CC_{(33)(nn)}^{(nn)}(l-1,p+1)= \CC_{(33)(nn)}^{(nn)}(l,p)}}
from the second one. In all these equations $x$, $y$, $a$ and $b$ are at values $(l,p)$. Note that $x^2+y^2=1$ (as well as $a^2+b^2=1$)  by normalization.

In order to solve these functional equations we use the fact that we know the value of the structure constants $\CC(1,p)$, i.e. the Virasoro ones. Also, by construction, the fields $\phi_{3,3}(l,p)$ and $\phi_{n,n}(l,p)$ are primary. Therefore their structure constants are just a product of lower level ones, as can be seen from the last of the equations \efor. One can then easily find:
\eqn\cnn{
\eqalign{
\CC_{(33)(nn)}^{(nn)}(l,p) &={\CG_n(p+l-1)\over\CG_n(p-1)},\cr
\CC_{(33)(nn)}^{(n+2n+2)}(l,p) &={\tilde\CG_n(p+l-1)\over\tilde\CG_n(p-1)}}
}
where we introduced the functions
\eqn\gamn{
\eqalign{
\CG_n(p)&=\[\g^3({p\ov p+1})\g^2({2\ov p+1})\g^2({n-1\ov p+1})\g^2({p-n\ov p+1})\g({3\ov p+1})\]^{1\ov 4},\cr
\tilde\CG_n(p)&=\[\g({p\ov p+1})\g({n\ov p+1})\g({p-n-1\ov p+1})\g({3\ov p+1})\]^{1\ov 4}. }}
and $\g(x)={\G(x)\ov \G(1-x)}$.

Finally, one can use the knowledge of the solutions for $l=1,2,4$ \refs{\df ,\zp ,\pogtri}. With all this, we can make a guess and check it directly. We present here only the constants we need in the sequel:
\eqn\resn{
\eqalign{
\CC_{(13)(nn)}^{(nn)}(l,p) &=-(n-1)\sqrt{{l\ov (p+l-2)(p-2)}} \CG_n(p+l-1),\cr
\CC_{(13)(nn)}^{(nn+2)}(l,p) &=\sqrt{{(p+l-2)(p-n-1)\ov (p+l-n-1)(p-2)}} \tilde\CG_n(p+l-1),\cr
\CC_{(13)(nn+2)}^{(nn+2)}(l,p) &=\(-l(n+1)+{2(p+l-2)(p-n-1)\ov p+l-n-1}\){\CG_{n+2}(p+l-1)\ov \sqrt{l(p+l-2)(p-2)}}.}}
We want to stress that the "structure constants" thus obtained are actually square roots of the true structure constants $C$. The reason is that our construction makes use of "chiral" one-dimensional fields instead of the real two-dimensional ones \myt.
Therefore the true structure constants are squares of those in \cnn\ and \resn.

The coefficients in the construction \defn\ are given by:
$$
x=\sqrt{{(l-1)(p-n-1)\ov l(p-n)}} \qquad y=\sqrt{{p+l-n-1\ov l(p-n)}}.
$$

In exactly the same way one obtains the structure constants involving the field $\tilde\phi_{n,n-2}(l,p)$. It turns out that they are obtained from the corresponding constants for
$\tilde\phi_{n,n+2}(l,p)$ by simply changing $n\rightarrow -n$:
\eqn\min{
\eqalign{
\CC_{(13)(nn)}^{(nn-2)}(l,p) &=\sqrt{{(p+l-2)(p+n-1)\ov (p+l+n-1)(p-2)}} \tilde\CG_{-n}(p+l-1),\cr
\CC_{(13)(nn-2)}^{(nn-2)}(l,p) &=\(l(n-1)+{2(p+l-2)(p+n-1)\ov p+l+n-1}\){\CG_{-n+2}(p+l-1)\ov \sqrt{l(p+l-2)(p-2)}}.}}


Finally, $\CC_{(13)(nn+2)}^{(nn-2)}(l,p)=0$ as can be seen by examining recursively the OPEs and fusion rules of the fields.

\newsec{Mixing of the fields}

Now we are in a position to describe the mixing of the fields we defined above along the RG flow.

We use here the renormalization scheme of \pogt\ . It
is a variation of that originally proposed by Zamolodchikov \zam\ .
The renormalized fields are expressed through the bare ones by:
$$
\phi^g_\a=B_{\a\b}(\l)\phi_\b
$$
(here $\phi$ could be any of the fields in consideration).
The two-point functions of the renormalized fields
\eqn\norm{
G_{\a\b}^g(x)=<\phi_\a^g(x)\phi_\b^g(0)>,\quad G_{\a\b}^g(1)=\delta_{\a\b} }
satisfy the Callan-Symanzik  equation
$$
(x\p_x-\b(g)\p_g)G_{\a\b}^g+\sum_{\rho=1}^2(\G_{\a\rho}G_{\rho\b}^g+\G_{\b\rho}G_{\a\rho}^g)=0
$$
where the matrix of anomalous dimensions $\Gamma$ is given by
\eqn\ano{ \G=B\hat\D
B^{-1}-\e\l B\p_\l B^{-1} }
where $\hat\D=diag(\D_1,\D_2)$ is a
diagonal matrix of the bare dimensions.
The matrix $B$ itself is
computed from the matrix of the bare two-point functions
using the normalization condition \norm\ and requiring the matrix
$\G$ to be symmetric.

Let us first consider the field $\phi_{n,n}$ defined by \defnn. It has a dimension close to zero and doesn't mix with other fields. This significantly simplifies \ano. Namely, for the anomalous dimension we simply obtain ($\phi_{n,n}^g=B(\l)\phi_{n,n}$):
$$
\D_{n,n}^{(g)}= \D_{n,n}+\e\l\p_\l \log B.
$$
Using the structure constant \resn\ obtained above we find (in terms of the renormalized coupling constant):
$$
\D_{n,n}^{(g)}= \D_{n,n}+{l(n^2-1)\e^2\pi g\ov 8\sqrt{3}}+O(g^2).
$$
So, at the fixed point
$$
\D_{n,n}^{(g^*)}={(n^2-1)l\e^2(2+3l\e+O(\e^2))\ov 32}
$$
which coincides up to this order with the dimension $\D_{n,n}(l,p-l)$ of the field $\phi_{n,n}(l,p-l)$ from \dimn. Thus, we conclude that, under the RG flow, the UV field $\phi_{n,n}(l,p)$ flows to the IR $\phi_{n,n}(l,p-l)$.

Now we turn to the mixing of the other fields in consideration.
Let us denote for convenience the basis of fields:
$$
\eqalign{ \phi_1=\tphi_{n,n+2},\quad
\phi_2=(2\D_{n,n}(2\D_{n,n}+1))^{-1}\p\bar\p \phi_{n,n},\quad
\phi_3=\tphi_{n,n-2} }
$$
where we normalized the field $\phi_2$ so that its bare two-point function is $1$. It is straightforward to
modify the functions involving $\phi_2$ taking into account the derivatives and the normalization.

The matrix of the two-point functions up to the first
order in the perturbation expansion was presented in \pert. It is obviously symmetric and expressed in terms of the structure constants as shown in \firstint.

Collecting all the dimensions and structure constants, we get in the leading order:

$$
\eqalign{
C^{(1)}_{1,1}&={4 (n+3)  \pi\ov l\sqrt 3 \e (n+1)},\quad
C^{(1)}_{1,2}=-{16  \sqrt{{n+2\ov n}} \pi\ov l\sqrt 3 \e (n+1) (n+3)},\quad
C^{(1)}_{1,3}=0,\cr
C^{(1)}_{2,2}&={16  \pi\ov l\sqrt 3  (n^2-1) \e} ,\quad
C^{(1)}_{2,3}=-{16  \sqrt{{n-2\ov n}} \pi\ov l\sqrt 3 \e (n-3) (n-1)},\cr
C^{(1)}_{3,3}&={4 (n-3)  \pi\ov l\sqrt 3 \e (n-1)}.}
$$

Now we can apply the renormalization procedure of \pogt\ and obtain
the matrix of anomalous dimensions \ano. The results, in terms of the renormalized coupling constant $g$, are:

$$
\eqalign{
\G_{1,1}&=\D_1+{2(n+3)  \pi g\ov l\sqrt 3 (n+1)},\quad
\G_{1,2}=\G_{2,1}=2{ (n-1) \sqrt{{n+2\ov 3n}} \pi g\ov l( n+1)}\cr
\G_{1,3}&=\G_{3,1}=0,\qquad
\G_{2,2}=\D_2+{8\sqrt 3 \pi  g\ov
 3 l (n^2-1)}\cr
\G_{2,3}&=\G_{3,2}={2 \sqrt{{n-2\ov 3n}} (n+1) \pi g\ov l(n-1)},\quad
 \G_{3,3}=\D_3+{2(n-3) \pi g\ov l\sqrt 3 (n-1)} }
 $$
where the dimensions are given by \dimn.

Evaluating this matrix at the fixed point \fx, we get
$$
\eqalign{
\G_{1,1}^{g^*}&=1 + {(20 - 4 n^2) \e\ov 8 (1 + n)}, \quad
\G_{1,2}^{g^*}=\G_{2,1}^{g^*}={(n-1) \sqrt{{n+2\ov n}} \e\ov n+1}\cr
\G_{1,3}^{g^*}&=\G_{3,1}^{g^*}=0,\qquad
\G_{2,2}^{g^*}=1 + {4 \e\ov n^2-1}\cr
\G_{2,3}^{g^*}&=\G_{3,2}^{g^*}={\sqrt{{n-2\ov n}} (n+1) \e\ov n-1},\quad
\G_{3,3}^{g^*}
=1 + {(n^2-5) \e\ov 2 (n-1)}  .}
$$
The eigenvalues of this matrix are:
$$
\eqalign{
\D_1^{g^*}&=1 +  {1 + n\ov 2} \e+O(\e^2) \cr
\D_2^{g^*}&=1+O(\e^2) \cr
\D_3^{g^*}&=1 + {1-n\ov 2} \e+O(\e^2)  . }
$$
This result coincides with dimensions $\tilde\D_{n+2,n}(l,p-l)$,$\D_{n,n}(l,p-l)+1$ and $\tilde\D_{n-2,n}(l,p-l)$ of the model $M(k-l,l)$ up to this order\foot{We note that this is true also in the next order where also $l$-dependence appear. The result will be presented elsewhere.}.

The corresponding normalized eigenvectors should be identified with the fields of $M(k-l,l)$:
\eqn\mix{
\eqalign{ \tphi_{n+2,n}^{(p-l)}&={2 \ov n (n+1)}\phi_1^{g^*} + {2
\sqrt{{n+2\ov n}}\ov n+1}\phi_2^{g^*} + {\sqrt{n^2-4}\ov
n}\phi_3^{g^*}\cr
\phi_2^{(p-l)}&={2 \sqrt{{n+2\ov n}}\ov n+1}\phi_1^{g^*} +{n^2-5\ov n^2-1}\phi_2^{g^*} -{2\sqrt{{n-2\ov n}}\ov n-1}\phi_3^{g^*}\cr
\tphi_{n-2,n}^{(p-l)}&={\sqrt{n^2-4}\ov n}\phi_1^{g^*}  - { 2
\sqrt{{n-2\ov n}}\ov n-1}\phi_2^{g^*} +{ 2\ov n(n-1)}\phi_3^{g^*}. }}
We used as before the notation $\tphi$ for the fields defined as in \defn\  and:
$$
\phi_2^{(p-l)}={1\ov 2\D_{n,n}^{(p-l)}(2\D_{n,n}^{(p-l)}+1)}\p\bar\p
\phi_{n,n}^{(p-l)}
$$
is the normalized derivative of the corresponding primary field. We notice that these eigenvectors are finite
as $\e\rightarrow 0$ with exactly the same entries as in $l=1$ and $l=2$ models.

\newsec{RG domain wall}

In the previous sections we proved that the coset CFT $M(k,l)$ perturbed by the field $\tilde\phi_{1,3}$ has a nontrivial fixed point corresponding to $M(k-l,l)$ in the leading order. We also found the mixing coefficients for certain fields between the UV $\CT_{UV}=M(k,l)$ and the IR $\CT_{IR}=M(k-l,l)$ theories.

Few years ago Gaiotto constructed a nontrivial conformal interface (RG domain wall) encoding the UV-IR map resulting through the RG flow described above \gai\ . Let us briefly recall the construction. Gaiotto considered a theory consisting of a IR $M(k-l,l)$ theory in the upper half plain and a UV $M(k,l)$ in the lower one. The conformal interface between the two CFT models is equivalent to some conformal boundary for the direct product of the theories $\CT_{UV}\times \CT_{IR}$:
$$
{\hat{su}(2)_k\times \hat{su}(2)_l\ov  \hat{su}(2)_{k+l}}\times {\hat{su}(2)_{k-l}\times \hat{su}(2)_l\ov  \hat{su}(2)_{k}}\sim {\hat{su}(2)_{k-l}\times \hat{su}(2)_l\times \hat{su}(2)_l\ov  \hat{su}(2)_{k+l}}.
$$
Note that two factors of $\hat{su}(2)_l$ appear at the RHS and therefore the theory possesses a natural $Z_2$ symmetry. In \gai\ it was shown that the desired boundary of the theory:
$$
\CT_B= {\hat{su}(2)_{k-l}\times \hat{su}(2)_l\times \hat{su}(2)_l\ov  \hat{su}(2)_{k+l}}
$$
acts as a $Z_2$ twisting mirror. Explicitly, this RG boundary is given by:
$$
|\tilde B>=\sum_{s,t}\sqrt{S^{(k-l)}_{1,t}S^{(k+l)}_{1,s}}\sum_d|t,d,d,s;\CB,Z_2\gg
$$
where the indices $t,d,s$ of the Ishibashi states refer to the representations of $\hat{su}(2)_{k-l}$, $\hat{su}(2)_{l}$,$\hat{su}(2)_{k+l}$ respectively and $S^{(k)}_{n,m}$ are the modular matrices of the $\hat{su}(2)_k$ WZNW model:
$$
S^{(k)}_{n,m}=\sqrt{{2\ov k+2}}\sin {\pi nm\ov k+2}.
$$
In this construction, the coefficients \mix\ of the UV-IR map are expressed in terms of the one point functions of the theory $\CT_{UV}\times \CT_{IR}$ in the presence of the RG boundary. So we need the explicit expression of the states corresponding to the fields $\phi^{IR}\phi^{UV}$ in terms of the states of the coset theory $\CT_B$.

Basic ingredient of the latter is the $\hat{su}(2)_k$ WZNW with a current $J$. As we mentioned above it is a CFT with central charge $c_k={3k\ov k+2}$. The primary fields $\phi_{j,m}$ and the corresponding states $|j,m>$ are labeled by the (half)integer spin $j$ and its projection $m=-j,-j+1,...,j$. Their conformal dimensions are given by:
\eqn\dimj{
\D_j={j(j+1)\ov k+2}.}
The representations are defined by the action of the currents on these states:
\eqn\act{
\eqalign{
J_0^{\pm}|j,m>&=\sqrt{j(j+1)-m(m\pm 1)}|j,m\pm 1>,\cr
J_0^{0}|j,m>&=m|j,m>.}}

Following \ppt\ let us denote by $K(z)$ and $\tilde K(z)$ the WZNW currents of $\hat{su}(2)_l$ entering the cosets of the IR and UV theories respectively. We reserve the notion $J(z)$ for the current of $\hat{su}(2)_{k-l}$ entering the IR coset. The corresponding energy momentum tensors can be expressed in terms of these currents using \suga. For example we can write symbolically the IR stress tensor as:
\eqn\tir{
T_{ir}={1\ov k-l+2}J^2+{1\ov l+2}K^2-{1\ov k+2}(J+K)^2}
and similarly for the UV one. Finally, we impose the condition that the state of the coset $\CT_B$ be a highest weight state of the diagonal current $J+K+\tilde K$.

Now we are in a position to compare the mixing coefficients in \mix\ with the corresponding one-point functions of the domain wall construction. Actually, we found it easier to compute the one-point functions of the other components of the corresponding multiplets. Namely, we shall consider the mixing of the "first components" given by the primary fields $\phi_{n,n\pm 2}$ and the first descendent of $\phi_{n,n}$ with respect to the current $A(z)$. Indeed, since $\phi_{n,n}$ belongs to the "vacuum sector" the current $A(z)$ is not branched around it and the dimension of the descendent $\tilde\phi_{n,n}=A_{-{2\ov l+2}}\phi_{n,n}$ is:
$$
\tilde\D_{n,n}={2\ov l+2}+{n^2-1\ov 4p}-{n^2-1\ov 4(p+l)}.
$$
So all these fields have dimension close to ${2\ov l+2}$ in the limit $p\rightarrow\infty$. Suppose they mix in the same way like it was in the case $l=2$ for example \ppt.
We want to compare the corresponding one point functions  with the coefficients in \mix.

We shall need therefore the explicit construction for the current $A(z)$. It goes in a way very similar to that of \ppt\ (see also \agt\ ). Consider for example the IR model. As in \ppt\ we take
\eqn\cur{
A(z)=C_a J^a(z)\phi_{1,-a}(z)+D_a K^a_{-1}\phi_{1,-a}(z)}
where $\phi_{1,m}(z)$ is a spin 1 field of the level $l$ WZNW theory with a current $K(z)$ and there is a summation over the index $a=\pm 1,0$. Indeed, the dimension of this current is:
$$
\D_A=1+{2\ov l+2}={l+4\ov l+2}.
$$
The coefficients $C_a$, $D_a$ are fixed by the requirement that the respective state be the highest weight state of the diagonal current algebra $J+K$. We get:
\eqn\cocur{
\eqalign{
D_+&={\k\over\sqrt{2}},\qquad D_0=\k,\qquad D_-=-{\k\over\sqrt{2}},\cr
C_+&=-\k {l+4\ov (k-l)\sqrt{2}},\quad C_0=-\k {l+4\ov (k-l)},\quad C_-=\k {l+4\ov (k-l)\sqrt{2}}}}
where $\k$ is a normalization constant. Since below we shall normalize the corresponding states we don't need it explicitly here.
It is straightforward to make a similar construction for the UV coset with obvious change of currents and levels.

Now we can pass to the computation of the one-point functions of the fields $\phi^{ir}\phi^{uv}$ and compare them with the corresponding coefficients in \mix.

Let us first start though with the field $\phi_{n,n}^{uv}$ itself. As we showed above it flows to the field $\phi_{n,n}^{ir}$ in the infrared. So we need to find the state in $\CT_B$ corresponding to $\phi_{n,n}^{ir}\phi_{n,n}^{uv}$. For this we need to match their conformal dimensions and to ensure that the state is a highest weight state of the diagonal current $J+K+\tilde K$. The dimension of the primary field $\phi_{n,n}$ can be read from \dimn. For the product of the IR and UV fields we have:
\eqn\nnd{
\D_{n,n}^{ir}+\D_{n,n}^{uv}={n^2-1\ov 4(k-l+2)}-{n^2-1\ov 4(k+l+2)}.}
It is easy to identify the corresponding state with
$$
|{n-1\ov 2},{n-1\ov 2}>|0,0>|0,0>
$$
where the three states correspond to $\hat{su}(2)$ of levels $k-l$ (with current $J$), IR level $l$ (with current $K$) and UV level $l$ (with current $\tilde K$) respectively. Indeed, this state is obviously a spin ${n-1\ov 2}$  highest weight state of $J+K+\tilde K$ and its dimension:
$$
\D_{{n-1\ov 2}}^J+\D_0^K+\D_0^{\tilde K}-\D_{{n-1\ov 2}}^{J+K+\tilde K}
$$
coincides with \nnd. It is obvious that this state is invariant under the $Z_2$ action, i.e. the exchange of the second and third factors. So the overlap of this state with its $Z_2$ image is just equal to $1$ and therefore:
\eqn\nnf{
<\phi_{n,n}^{ir}\phi_{n,n}^{uv}|RG>={\sqrt{S_{1,n}^{(k-l)}S_{1,n}^{(k+l)}}\ov S_{1,n}^{(k)}}=1+{3l^2\ov 4k^2}+O({1\ov k^3}).}
This confirms that up to the leading order in $k\rightarrow\infty$ the field  $\phi_{n,n}^{uv}$ flows to $\phi_{n,n}^{ir}$.

Note that the calculations are very similar to that of \ppt\ . We shall see that this is the case also for the other one-point functions below.

Let us find for example the state corresponding to $\phi_{n+2,n}^{ir}\phi_{n,n+2}^{uv}$. The dimensions can be found from \dimn\ and we have:
$$
\D_{n+2,n}^{ir}+\D_{n,n+2}^{uv}={4\ov l+2}+{(n+1)(n+3)\ov 4(k-l+2)}-{(n+1)(n+3)\ov 4(k+l+2)}.
$$
In exactly the same way as in \ppt\ the corresponding state should have the form:
\eqn\pp{
\sum_{\a,\b=\pm 1,0}C_{\a\b}|{n+1\ov 2},{n+1\ov 2}-\a-\b>|1,\a>|1,\b>.}
The coefficients $C_{\a\b}$ are obtained by imposing the condition that \pp\ has a correct IR dimension and is a highest weight state of $J+K+\tilde K$. We obtain:
$$
C_{++}=-{1\ov \sqrt{n}}C_{0+},\qquad C_{-+}=-{\sqrt{{n+1\ov 2}}}C_{0+}
$$
and all the other coefficients vanish. The overall normalization fixes
$$
C_{0+}^2={2n\ov (n+1)(n+2)}.
$$
Taking the overlap of the state \pp with its $Z_2$ image we find:
\eqn\ppf{
<\phi_{n+2,n}^{ir}\phi_{n,n+2}^{uv}|RG>={2\ov (n+1)(n+2)}{\sqrt{S_{1,n+2}^{(k-l)}S_{1,n+2}^{(k+l)}}\ov S_{1,n}^{(k)}}={2\ov n(n+1)}+O({1\ov k^2}).}

Again, we see that the calculations are very similar to those of \ppt\ . The reason is that the spins entering the construction of the state \pp\ are the same although the levels of $\hat{su}(2)$ algebras are different. The latter fact however doesn't affect the final result. This is valid also for the other calculations involving the primary fields so we present here just the result:
\eqn\others{
\eqalign{
<\phi_{n-2,n}^{ir}\phi_{n,n-2}^{uv}|RG>&={2\ov (n-1)(n-2)}{\sqrt{S_{1,n-2}^{(k-l)}S_{1,n-2}^{(k+l)}}\ov S_{1,n}^{(k)}}={2\ov n(n-1)}+O({1\ov k^2}),\cr
<\phi_{n+2,n}^{ir}\phi_{n,n-2}^{uv}|RG>&={\sqrt{S_{1,n+2}^{(k-l)}S_{1,n-2}^{(k+l)}}\ov S_{1,n}^{(k)}}={\sqrt{n^2-4}\ov n}+O({1\ov k^2}),\cr
<\phi_{n-2,n}^{ir}\phi_{n,n+2}^{uv}|RG>&={\sqrt{S_{1,n-2}^{(k-l)}S_{1,n+2}^{(k+l)}}\ov S_{1,n}^{(k)}}={\sqrt{n^2-4}\ov n}+O({1\ov k^2}).}}

Consider in more details the functions involving the descendent field $\tilde\phi_{n,n}$. Let us first consider $\phi_{n,n}^{ir}\phi_{n,n+2}^{uv}$:
\eqn\desd{
\D_{n,n}^{ir}+\D_{n,n+2}^{uv}={2\ov l+2}+{n^2-1\ov 4(k-l+2)}-{(n+1)(n+3)\ov 4(k+l+2)}.}
The corresponding state is:
\eqn\stat{
|{n-1\ov 2},{n-1\ov 2}>|0,0>|1,1>}
(because the spin $1$ term in \desd\ refers to UV level $l$ current $\tilde K$). Using the explicit expression of the current it is easy to find that for the descendent we have:
$$
\eqalign{
A_{-{2\ov l+2}}|{n-1\ov 2},{n-1\ov 2}>|0,0>|1,1>&=C_aJ_0^a|{n-1\ov 2},{n-1\ov 2}>|1,-a>|1,1>+\cr
&+D_aK_0^a|{n-1\ov 2},{n-1\ov 2}>|1,-a>|1,1>}
$$
where the coefficients are given by \cocur. This gives:
$$
\eqalign{
A_{-{2\ov l+2}}|{n-1\ov 2},{n-1\ov 2}>|0,0>|1,1>&=\k {(l+4)\ov (k-l)}\sqrt{{n-1\ov 2}}|{n-1\ov 2},{n-3\ov 2}>|1,1>|1,1>-\cr
&-\k {(l+4)\ov (k-l)}{n-1\ov 2}|{n-1\ov 2},{n-1\ov 2}>|1,0>|1,1>.}
$$
The normalization condition is:
\eqn\norm{
{\k^2(l+4)^2\ov (k-l)^2}{n^2-1\ov 4}=1.}
Thus, for the one-point function we get:
\eqn\nnex{
<\tilde\phi_{n,n}^{ir}\phi_{n,n+2}^{uv}|RG>={2\ov n+1}{\sqrt{S_{1,n}^{(k-l)}S_{1,n+2}^{(k+l)}}\ov S_{1,n}^{(k)}}={2\ov n+1}\sqrt{{n+2\ov n}}+O({1\ov k^2}).}

It is clear that also the other  calculations for the one-point functions including the descendent $\tilde\phi_{n,n}$ will be similar to \ppt\ . The only difference appear in the level dependence of the coefficients and consequently in the normalization like \norm. Since the overlap is also quadratic this level dependence disappear. The only level dependence left is thus in the modular matrix. Finally, we get:
\eqn\nnrest{
\eqalign{
<\tilde\phi_{n,n}^{ir}\phi_{n,n-2}^{uv}|RG>&=-{2\ov n-1}{\sqrt{S_{1,n}^{(k-l)}S_{1,n-2}^{(k+l)}}\ov S_{1,n}^{(k)}}=-{2\ov n-1}\sqrt{{n-2\ov n}}+O({1\ov k^2}),\cr
<\phi_{n-2,n}^{ir}\tilde\phi_{n,n}^{uv}|RG>&=-{2\ov n-1}{\sqrt{S_{1,n-2}^{(k-l)}S_{1,n}^{(k+l)}}\ov S_{1,n}^{(k)}}=-{2\ov n-1}\sqrt{{n-2\ov n}}+O({1\ov k^2}),\cr
<\phi_{n+2,n}^{ir}\tilde\phi_{n,n}^{uv}|RG>&={2\ov n+1}{\sqrt{S_{1,n+2}^{(k-l)}S_{1,n}^{(k+l)}}\ov S_{1,n}^{(k)}}={2\ov n+1}\sqrt{{n+2\ov n}}+O({1\ov k^2}),\cr
<\tilde\phi_{n,n}^{ir}\tilde\phi_{n,n}^{uv}|RG>&={n^2-5\ov n^2-1}{\sqrt{S_{1,n}^{(k-l)}S_{1,n}^{(k+l)}}\ov S_{1,n}^{(k)}}={n^2-5\ov n^2-1}+O({1\ov k^2}).}}

We see that all these results \ppf,\others,\nnex\ and \nnrest\ are in a perfect agreement with the leading order calculations \mix\ presented in the previous section.

\newsec{Concluding remarks}

We presented the calculation of the mixing coefficients of certain fields in the general $\hat{su}(2)$ coset models $M(k,l)$ perturbed by the least relevant field. This was done using two different approaches. First we used a perturbation theory in $1/k$, $\k\rightarrow\infty$. We found the corresponding structure constants governing the perturbation expansion. This was done using the projected tensor product construction of the coset model \myt\ . In the second part of this paper we presented the RG domain wall construction of Gaiotto for the general coset model. We found the one-point functions expressing the corresponding mixing coefficients. It turns out that both results coincide in the leading order. Moreover, they are finite, do not depend on $l$ and coincide with the corresponding coefficients for $l=1$ and $l=2$. It is interesting to compare these two approaches in the next to leading order. We will present the necessary calculations elsewhere. Also, it will be good to find the results for other fields, for example the analogs of $\phi_{n,n\pm 1}$ in Virasoro and superconformal theories. Finally, an interesting problem is the  extension of these constructions to other two-dimensional CFT's.

\vskip1cm
\no{\bf Acknowledgements}

\no This work was supported in part by the Bulgarian NSF Grant DFNI T02/6.

\listrefs

\bye